\documentclass[aps,preprint]{revtex4}%
\usepackage{amsfonts}
\usepackage{amsmath}
\usepackage{amssymb}
\usepackage{graphicx}%
\providecommand{\U}[1]{\protect \rule{.1in}{.1in}}

\begin{document}
\preprint{ }
\title[Nonlocal optical model potential]{Solution of the Schr\"{o}dinger equation containing a Perey-Buck nonlocality.}
\author{George H. Rawitscher}
\affiliation{Physics Department, University of Connecticut, Storrs, CT 06269}
\pacs{PACS number}

\begin{abstract}
The solution of a radial Schr\"{o}dinger equation\ for $\psi(r)$ containing a
nonlocal potential of the form $\int K(r,r^{\prime})\  \psi(r^{\prime
})\ dr^{\prime}$ is obtained to high accuracy by$\ $means of two methods. An
application to the Perey-Buck nonlocality is presented, without using a local
equivalent representation. The first method consists in expanding $\psi$ in a
set of Chebyshev polynomials, and solving the matrix equation for the
expansion coefficients numerically. An accuracy of between $1:10^{-5}$ to
$1:10^{-11}$ is obtained, depending on the number of polynomials employed. The
second method consists in expanding $\psi$ into a set of $N_{S}$ Sturmian
functions of positive energy, supplemented by an iteration procedure. For
$N_{S}=15$ an accuracy of $1:10^{-4}$ is obtained without iterations. After
one iteration the accuracy is increased to $1:10^{-6}.$ Both methods are
applicable to a general nonlocality $K$. The spectral method is less complex
(requires less computing time) than the Sturmian method, but the latter can be
very useful for certain applications.

\end{abstract}
\startpage{1}
\endpage{102}
\maketitle

\section{\bigskip Introduction}

The solution of the Schr\"{o}dinger equation in the presence of a nonlocal
potential has been the subject of many investigations since 1934
\cite{Watagh}.\ In the optical model one source of nonlocality occurs in order
to describe the knock-on scattering (a manifestation of the identity of
nucleons and of the Pauli exclusion principle). Another nonlocality occurs in
order to describe the dynamic polarization of the target or projectile during
the scattering process (The Feshbach potential). An interesting study of the
relation between both nonlocalities and the microscopic structure of target
nuclei has recently been presented \cite{AMOS}. The knock-on process leads to
a semi-separable rank one nonlocality (the meaning of "rank" is explained in d
Appendix A), while the dynamic polarization nonlocality leads to a general
kernel $K(\vec{r},\vec{r}^{\  \prime})$ that acts on the wave function $\psi$
in the form of an integral in the Schr\"{o}dinger equation%
\begin{equation}
\left[  -\nabla^{2}+V(\vec{r})-k^{2}\right]  \psi(\vec{r}^{\  \prime})=-\int
K(\vec{r},\vec{r}^{\  \prime})\psi(\vec{r}^{\prime})d^{3}\vec{r}^{\  \prime}.
\label{SCHR3}%
\end{equation}
The potential $V$ and the integral over the kernel $K$ are in units of inverse
length squared, and are obtained by transforming them from their energy units
into $fm^{-2}$ units by multiplication by the well known factor $2m/\hslash
^{2}$. Here $m$ is the reduced mass of the incident particle, $\hslash$ is
Plank's constant divided by $2\pi$, $V$ is the local part of the potential
including the spin orbit interaction, and $k$ is the wave number of the
incident projectile, related to the center of mass energy $E$ by
$(2m/\hslash^{2})E=k^{2}.$ For application to the case of nucleon scattering
from a nucleus, $(\hslash^{2}/2m)$ can be set approximately equal to $20.4$
$MeV\ fm^{-2}.$ Other\ analytical forms of nonlocalities have also been
introduced \cite{Cooper}, and a particular velocity dependent form is
described in Ref. \cite{MAHM-1}.

For the case that $K$ is a semi-separable rank one operator, the solution of
Eq. (\ref{SCHR3}) was initially obtained by various laborious combinations of
solutions of a local equation \cite{SK}, subsequently a perturbative method
using the Singular Value Decomposition (SVD) \cite{SVD} was developed
\cite{ESSAID}, and later a solution using a spectral \cite{SPECTR} expansion
into Chebyshev polynomials \cite{IEM} was presented \cite{EXCH}. The latter
did not make use of perturbation theory but applied only to this particular
semi-separable rank one nonlocality. Other methods of solution for the
nonlocality of the type of Eq. (\ref{SCHR3}) have also been developed. In 1990
Kim and Udagawa \cite{UDAGAWA} presented an efficient solution for a general
nonlocal potential $K$ using a Lanczos \ iterative method. Other iterative
methods were also developed \cite{Sinha}, \cite{Mack}. Expansions into powers
of the momentum operator have been presented \cite{Baseia}, \cite{Lo}, and
local equivalent potentials have been developed \cite{Perey} and used widely
\cite{Hussein}. A different method of solution, employing expansions into
Chebyshev polynomials, has also been described previously \cite{KANG} for
nonlocal potentials.

It is the purpose of the present article to present an alternate method,
different from either method described above, by using a combination of the
Spectral method and the SVD method. The method is nonperturbative, does not
depend on a choice of an auxiliary local potential $U_{0}$, can reach an
accuracy better than $1:10^{-11},$ is applicable to a general kernel $K,$ and
its complexity is less than that of Ref. \cite{KANG}. A numerical example is
given for the case of the Perey-Buck nonlocality \cite{Perey}, and a
comparison with an iterative expansion into Sturmian functions \cite{STURM-GR}
is also be presented. That comparison not only serves as a check on the
Spectral method presented here, but also serves to underline the usefulness of
Sturmian expansions. As an example, Sturmian functions were found
\cite{Canton} to be very instrumental in describing resonances in low energy
nucleon-nucleus scattering, as well as in implementing the Pauli exclusion
principle, both of which are not generally taken account of in the
conventional numerical solution of coupled channel equations.

In section II the version of the Perey-Buck nonlocality kernel used for the
numerical application will be defined, in Section III the spectral method
combined with the SVD decomposition will be described, in Section IV a
sturmian expansion method for the wave function solution, together with an
iterative correction method, is described, and Section V contains the summary
and conclusions.

\section{The Perey-Buck nonlocality}

\begin{figure}
[ptb]
\begin{center}
\includegraphics[
height=2.1655in,
width=2.8807in
]%
{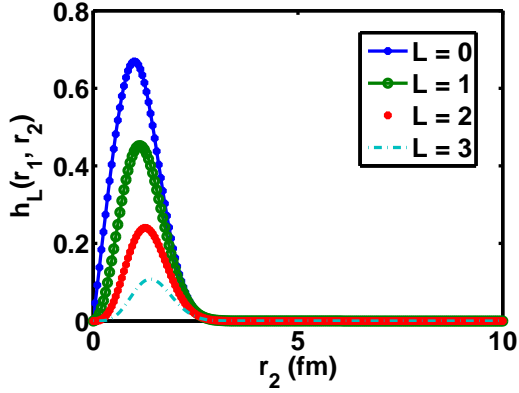}%
\caption{The Perey-Buck nonlocality function $h_{L}(r_{1},r_{2})$.for
$r_{1}=1.0\ fm$, for various values of the partial wave angular momentum $L.$}%
\label{FIG1}%
\end{center}
\end{figure}

\begin{figure}
[ptb]
\begin{center}
\includegraphics[
height=2.2355in,
width=2.9715in
]%
{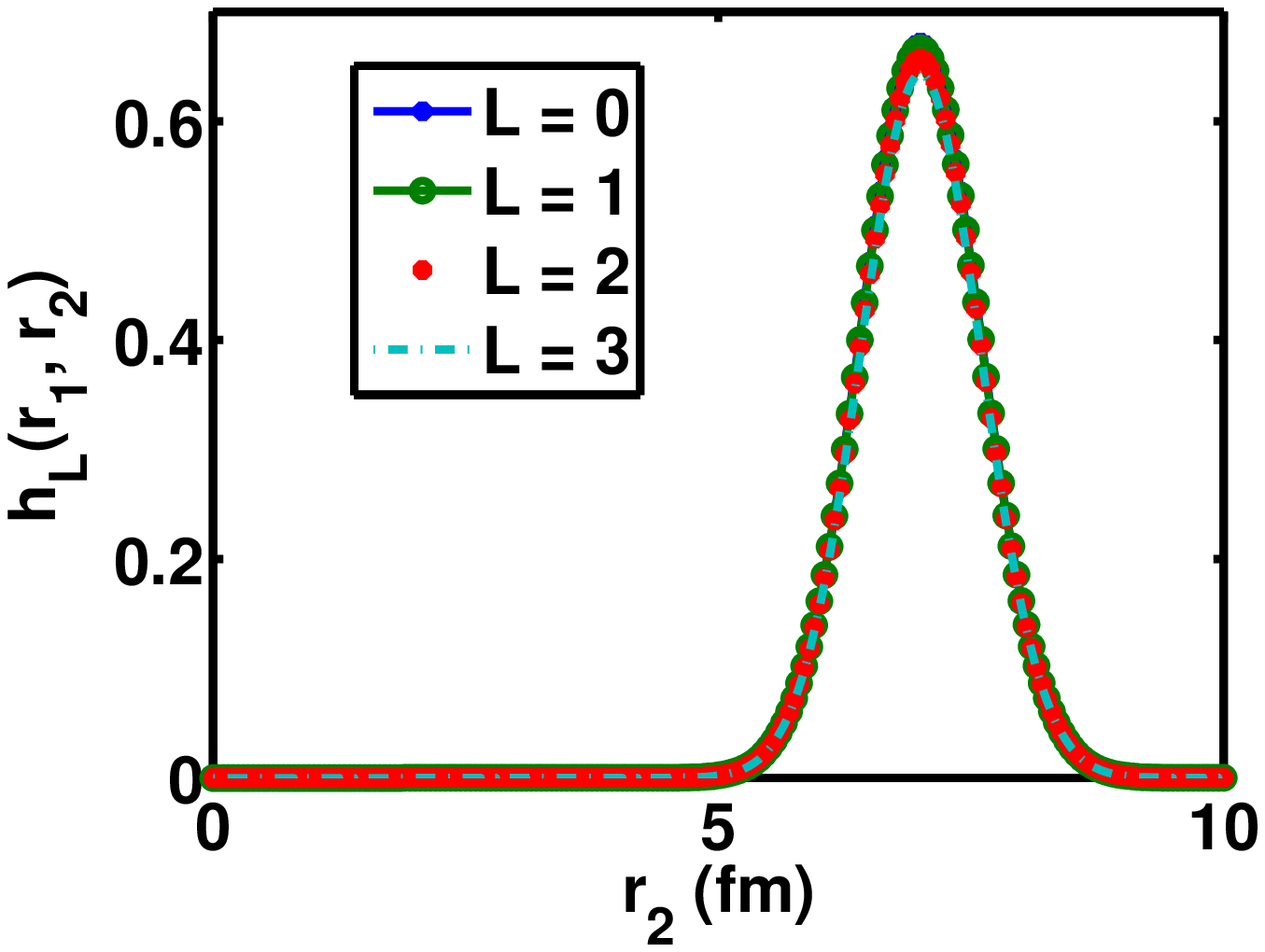}%
\caption{Same as Fig. \ref{FIG1} for $r_{1}=7fm.$}%
\label{FIG2}%
\end{center}
\end{figure}

This nonlocality was first introduced by Frahn and Lemmer \cite{FRAHN}, and
developed further by Perey and Buck \cite{Perey}. The kernel is of the form%
\begin{equation}
\bar{K}(\vec{r},\vec{r}^{\prime})=U(\frac{1}{2}|\vec{r}+\vec{r}^{\prime
}|)\  \frac{1}{\pi^{3/2}\  \beta^{3}}e^{-\ [\vec{r}-\vec{r}^{\prime}]^{2}%
/\beta^{2}}, \label{K-GEN}%
\end{equation}
where $\beta$ is the nonlocality parameter. For the present application a
simpler form will be adopted,%
\begin{equation}
K(\vec{r},\vec{r}^{\prime})=V_{PB}(r)\ h(\vec{r},\vec{r}^{\prime})
\label{K-SIMP}%
\end{equation}
with
\begin{equation}
h(\vec{r},\vec{r}^{\prime})=\frac{1}{\pi^{3/2}\  \beta^{3}}e^{-\ [\vec{r}%
-\vec{r}^{\prime}]^{2}/\beta^{2}}. \label{h}%
\end{equation}
The kernel $K$ given by Eq. (\ref{K-SIMP}) is convenient for the purpose of
numerical calculation but is\ no longer symmetric. For the partial wave
decomposition, for each value of the angular momentum number $L$ the angular
momentum projection $h_{L}(r,r^{\prime})$ of the operator $h$ is given in the
Appendix of Ref. \cite{Perey} in terms of Spherical Bessel functions and of
gaussian exponentials in $r$ and $r^{\prime}$. The behavior of $h_{L}%
(r,r^{\prime})$ as a function of $r^{\prime}$ (indicated as $r_{2})$ is
illustrated in Figs. \ref{FIG1} and \ref{FIG2} for two values of $r$
(indicated as $r_{1}),$ respectively, and using for $\beta$ the standard value
$0.84\ fm.$%

%

\begin{figure}
[ptb]
\begin{center}
\includegraphics[
height=2.1248in,
width=2.8253in
]%
{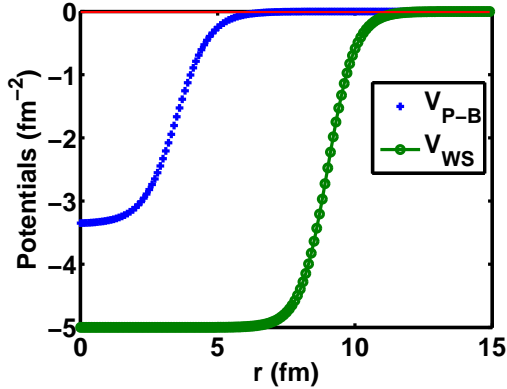}%
\caption{Potentials $V_{PB}$ and $V_{WS}$ as a function of $r.$ Both
potentials are of the Woods Saxon type, given by Eq. (\ref{VPB}). $V_{PB}$ is
the potential that is multiplied into the nonlocal term $h_{L}(r,r^{\prime})$
\ according to Eq. (\ref{K-SIMP})\ with parameters given in Eq. (\ref{PAR_PB}%
). The one denoted as $V_{WS}$ has parameters given by $(V_{0},R,a)=(-5fm^{-2}%
,9fm,0.5fm),$ and is used to define the Sturmian functions described further
below. }%
\label{FIG3}%
\end{center}
\end{figure}

One sees from these figures that the angular momentum dependence of
$h_{L}(r,r^{\prime})$ is reduced with increasing values of $r$, and that
$h_{L}$ peaks in the vicinity of $r.$ The decrease of $K_{L}(r,r^{\prime
})=V_{PB}(r)\ h_{L}(r,r^{\prime})$ with distance $r$ or $r^{\prime}$ is
assured by the decrease with $r$ of the factor $V_{PB}(r)$. The latter is
taken to be of the Woods-Saxon form%
\begin{equation}
V_{PB}(r)=V_{0}/(1+\exp[(r-R)/a]),\label{VPB}%
\end{equation}
using for the parameters the values%
\begin{equation}
V_{0}=-3.36\ fm^{-2},\ R=3.5\ fm,\ a=0.6\ fm\label{PAR_PB}%
\end{equation}
This potential is illustrated in Fig. \ref{FIG3}, where it is also compared
with a potential $V_{WS}$ which will be used to generate a set of Sturmian
basis functions, to be used in a subsequent section.%

\section{The Spectral Method}

A version of the spectral method employed here was developed recently
\cite{IEM}. It consists in dividing the radial interval into partitions, and
obtaining two independent solutions of the Schr\"{o}dinger Eq. (\ref{SCHR3})
in each partition. These solutions are obtained by transforming Eq.
(\ref{SCHR3}) into an equivalent Lippmann-Schwinger integral equation (L-S),
and solving the latter by expanding the solution into Chebyshev functions,
mapped to the interval $[-1,+1]$. The corresponding discretized matrices are
not sparse, but are of small dimension, and the two independent functions are
very precise (accuracy of $1:10^{-11})$. The solution $\psi$ in each partition
is obtained by a linear combination of the two independent functions, and the
matrix required to obtain the expansion coefficients has a dimension twice as
large as the number of partitions, but it is sparse. Details are given in Ref.
\cite{IEM}, and a pedagogical version is found in Ref. \cite{STRING}. For the
present application the division \ of the radial interval into partitions is
not made, because the effect of the nonlocal potential would extend into more
than one partition, making the programming more cumbersome.

In the presence of a nonlocal potential $K$, the (L-S) equation for the
partial wave function $\psi$ and angular momentum number $L=0$ takes the form
\begin{equation}
\psi(r)=F(r)+\int_{0}^{\infty}\mathcal{G}_{0}\mathcal{(}r,r^{\prime
})[V(r^{\prime})\delta(r^{\prime}-r^{\prime \prime})+K(r^{\prime}%
,r^{\prime \prime})]\psi(r^{\prime \prime})\ dr^{\prime \prime}dr^{\prime},
\label{L-S}%
\end{equation}
where $\mathcal{G}_{0}\mathcal{(}r,r^{\prime})$ is the Green's function given
by
\begin{equation}
\mathcal{G}_{0}(r,r^{\prime})=-\frac{1}{k}F(r_{<})\times H(r_{>}), \label{L6}%
\end{equation}
where $(r,r\prime)=(r_{<},r_{>})$ if $r\leq r^{\prime}$ and $(r,r\prime
)=(r_{>},r_{<})$ if $r\geq r^{\prime}$, where
\begin{equation}
F(r)=\sin(kr);~~~H(r)=\cos(kr)+i\sin(kr), \label{L7}%
\end{equation}
and where $k$ is the wave number.

The nonlocal part $K$ \ of the potential is expanded by means of the Singular
Value Decomposition (SVD) as follows. First a set of Chebyshev support points
$\xi_{i}$, $i=1,2,..N$ are defined in each partition and the corresponding
discretized matrix $\  \mathcal{K}_{i,j}=K(\xi_{i}$,$\xi_{j})$ is obtained,
where $N$ is the number of Chebyshev polynomials $T_{0},$ $T_{1},$
$...T_{N-1}$ to be used in the expansion. These support points are the zeros
of the first Chebyshev polynomial $T_{N}(x)$ not used in the expansion, mapped
into the radial partition interval, as described previously \cite{IEM,
STRING}. However, in the present application only one large partition is used.
The singular value decomposition of a $N\times N$ matrix $\mathcal{K}$ is
given by \cite{SVD}, $\mathcal{K=}U\boldsymbol{S}V^{\dag},$ or $\mathcal{K}%
_{i,j}\  \mathcal{=}\sum_{s=1}^{N}u_{i,s}\sigma_{s}(v_{j,s})^{\dag
},~~~i,j=1,2,..,N.$ This result can be expressed in terms of the column
vectors $\boldsymbol{u}_{s}$ and $\boldsymbol{v}_{s}$ of the $N\times N$
unitary matrices $U$ and $V$, respectively%
\begin{equation}
\mathcal{K}_{i,j}\  \mathcal{=}\sum_{s=1}^{N}\boldsymbol{u}_{s}(i)\sigma
_{s}[\boldsymbol{v}_{s}(j)]^{\dag}, \label{76}%
\end{equation}
and $\boldsymbol{S}$ is a diagonal matrix containing the singular values
$\sigma_{s}$, with $s=1,2,..,N.$ The $\sigma_{s}$ are positive numbers,
ordered in descending values, and the symbol $\dag$ signifies transposition
and complex conjugation. The matrices $U$ and $V^{\dag}$ are unitary, but are
not orthogonal to each other.

According to the Eq. (\ref{76}) the matrix $\mathcal{K}$ can be expressed in
terms of a sum of products $(\mathbf{u})\times(\boldsymbol{v})^{\dag}$, each
of which represents a rank one $N\times N$ matrix. Thus, if the values of the
wave function $\psi$ at the support points $\xi_{j}$\ were expressed as a
column vector $\boldsymbol{\psi}\emph{,}$ with $\boldsymbol{\psi}_{j}=\psi
(\xi_{j})$, then the integral in over $dr^{\prime \prime}$ of the
dot-product\ $[\boldsymbol{v}_{s}]^{\dag}\cdot \boldsymbol{\psi}$ , as required
in Eq. (\ref{L-S}),$\ $would be given by
\begin{equation}
\langle \boldsymbol{v}_{s}|\boldsymbol{\psi \rangle=}\sum_{j=1}^{N}v^{\dag
}(s,j)w_{j}\boldsymbol{\psi}_{j}, \label{L8}%
\end{equation}
where $w_{j}$ are integration weights. Hence the action of $K$ on $\psi$ can
be broken into simpler expressions of the type
\begin{equation}
\left(  \mathcal{K}\boldsymbol{\psi}\right)  _{i}\boldsymbol{=}\sum_{s=1}%
^{N}\  \boldsymbol{u}_{s}(i)\sigma_{s}\langle \boldsymbol{v}_{s}%
|\boldsymbol{\psi \rangle.} \label{L9}%
\end{equation}
In the spectral method, the integral $\langle \boldsymbol{v}_{s}%
|\boldsymbol{\psi \rangle}$ is done by mapping $\psi(r)\rightarrow \bar{\psi
}(x)$ onto the interval $[-1\leq x\leq+1],$ by expanding $\bar{\psi}(x)$ into
Chebyshev polynomials,
\begin{equation}
\bar{\psi}(x)=\sum_{j=1}^{N}\ a_{j}T_{j-1}(x). \label{L9a}%
\end{equation}
and by using the integral properties of the Chebyshev polynomials. The various
overlap integrals involving Chebyshev polynomials are obtained by a method
denoted as the Chebyshev-Gauss integration procedure, and is given by Eq.
($50$) of Ref. \cite{STRING}. By this means Eq. (\ref{L-S}) can be transformed
into an equation containing matrices acting on the expansion coefficients
$a_{j}$, and its algebraic solution permits one to obtain the coefficients
$a_{j}$, and hence $\psi(r)$ via Eq. (\ref{L9a}). The\ method utilizing the
SVD expansion, described above, can be utilized for any non-singular matrix,
and hence the method applies to any non-singular kernel $K.$

\begin{figure}
[ptb]
\begin{center}
\includegraphics[
height=2.2252in,
width=2.9568in
]%
{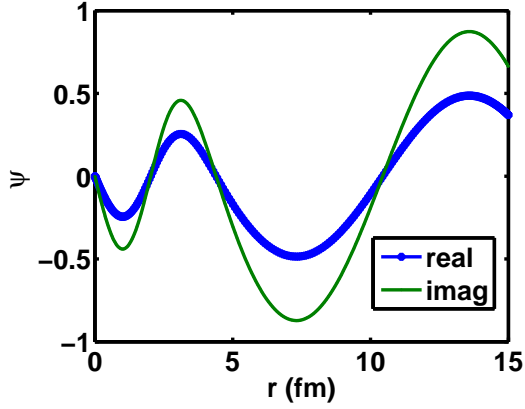}%
\caption{The wave function $\psi(r),$ solution of the Schroedinger eq.
(\ref{L-S}) with the nonlocal potential kernel $K$ given by Eq.(\ref{K-SIMP}).
The solution is obtained by the spectral Chebyshev expansion method for
$k=0.5fm^{-1}$}%
\label{FIG5}%
\end{center}
\end{figure}

\begin{figure}
[ptb]
\begin{center}
\includegraphics[
height=2.1776in,
width=2.8954in
]%
{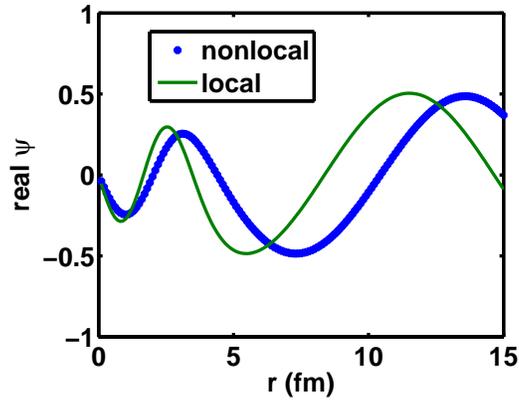}%
\caption{Comparison of the real parts of the local and nonlocal wave functions
$\psi$. The nonlocal one assumes the presence of the Perey-Buck nonlocality,
the local one replaces the nonlocality $h_{L}(r,r^{\prime})$ by $\delta
(r-r^{\prime})$. Here $h_{L}$ is defined in Eq. (\ref{h}) with $L=0$ and
$k=0.5\ fm^{-1}$}%
\label{FIG7}%
\end{center}
\end{figure}

\begin{figure}
[ptb]
\begin{center}
\includegraphics[
height=2.1664in,
width=2.8807in
]%
{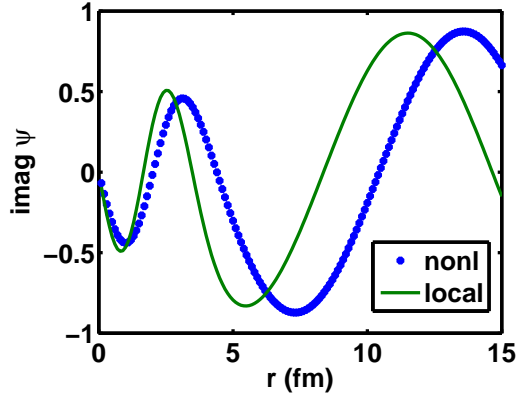}%
\caption{Same as Fig. \ref{FIG7} for the imaginary parts of the wave
functions.}%
\label{FIG7A}%
\end{center}
\end{figure}

The solution $\psi$ of Eq. (\ref{L-S}) for $k=0.5\ fm^{-1}$ is displayed in
Fig. \ref{FIG5}.
The effect of the nonlocality is to "push out" the wave function to larger
distances. This can be seen from Figs. \ref{FIG7} and \ref{FIG7A}, which
compare the real and imaginary parts of the solution $\psi$ for
$k=0.5\ fm^{-1}$ in the presence of the nonlocality with the solution for
which the non-locality, given by Eq. (\ref{h}), is replaced by a delta
function, $h_{L}(r,r\prime)\rightarrow \delta(r-r^{\prime})$.

\subsection{Accuracy analysis}

\begin{figure}
[ptb]
\begin{center}
\includegraphics[
height=2.2295in,
width=2.9646in
]%
{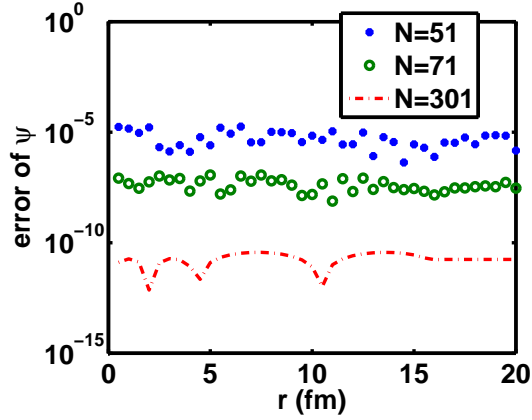}%
\caption{Estimate of the error of the spectral method, by comparison of the
wave function $\psi$ for various values $N$ with each other. Here $N$ is the
number of Chebyshev polynomials used in the solution of Eq. (\ref{L-S}) in the
radial interval $[0,20\ fm]$. Plotted are the absolute values of the
difference of $\psi$ between the results for $N=51,\ N=71$ and $N$ $=301$,
respectively, with the result for $N=$ $501$.}%
\label{FIG8}%
\end{center}
\end{figure}

The error of the spectral method for the nonlocal potential $K$ is obtained by
comparing solutions for three different values $N=51,71,$ and $301$ of the
number of Chebyshev expansion functions used in the radial interval
$[0,20\ fm]$ with the solution obtained with $N=501$. The results are shown in
Fig. \ref{FIG8}, and listed in Table \ref{TABLE1}, which also displays the
respective computing times.%
The calculations are done in MATLAB, on a desktop using an Intel TM2 Quad,
with a CPU Q 9950, a frequency of 2.83 GHz, and a RAM\ of 8 GB.
\begin{table}[tbp] \centering
\begin{tabular}
[c]{||l|l|l|}\hline
$N$ & $accuracy$ & $time(s)$\\ \hline \hline
$51$ & $2\times10^{-5}$ & $0.4$\\ \hline
$71$ & $10^{-7}$ & $0.7$\\ \hline
$301$ & $4\times10^{-11}$ & $13$\\ \hline
\end{tabular}
\caption{Accuracy and computing time of the spectal method. N is the number of Chebyshev support points in the radial interval [0,20 {\sl fm}]}\label{TABLE1}%
\end{table}
Had the radial interval been subdivided into partitions, a further decrease of
computing time would have been achieved. Such partition division is required
if the wave function extends out to large distances, as was the case, for
example, in the calculation of the Helium di-atom bound state \cite{HEHE} in
the presence of a local He-He potential.

\begin{figure}
[ptb]
\begin{center}
\includegraphics[
height=1.9259in,
width=2.5616in
]%
{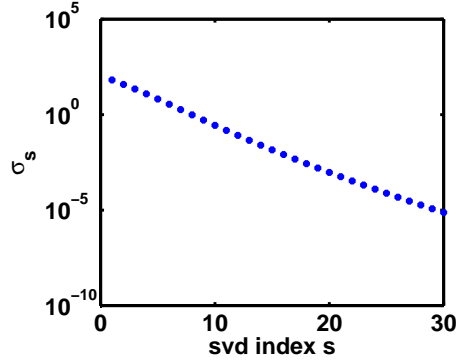}%
\caption{The singular values $\sigma_{s},\ s=1,2,..30$, for the Singular Value
decomposition of the nonlocality kernel $K_{L}(r,r^{\prime})=V_{PB}%
(r)\ h_{L}(r,r^{\prime})$, for $L=0$. The function $h_{L}$ is defined in Eq.
(A.2) in Ref. \cite{Perey} and is illustrated in Figs. \ref{FIG1} and
\ref{FIG2}, $\ $ $V_{PB}$ is defined by Eqs. (\ref{VPB}) and (\ref{PAR_PB}),
and is illustrated in Fig. \ref{FIG3}. The number of Chebyshev support points,
which is equal to the number of Chebyshev polynomials used, is $N=302.$}%
\label{FIG4}%
\end{center}
\end{figure}

\begin{figure}
[ptb]
\begin{center}
\includegraphics[
height=2.2234in,
width=2.9577in
]%
{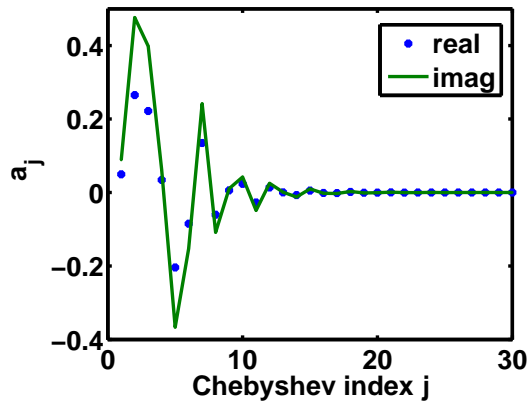}%
\caption{The Chebyshev expansion coefficients $a_{j}$ of the wave function
$\psi$ that satisfies the nonlocal Eq. (\ref{L-S}), where $K(r,r^{\prime})$ is
the Perey-Buck Kernel (\ref{K-SIMP}). The radial interval is $[0,15fm]$, the
wave number $k=0.5\ fm^{-1},$ and the total number $N$ of Chebyshev
polynomials is $302$.}%
\label{FIG6}%
\end{center}
\end{figure}

An illustration of the singular values $\sigma_{s}$ for the case of the
Perey-Buck nonlocal kernel, using only one partition from $r=0$ to $r_{\max
}=15,$ with $N=301$, is shown in Fig. \ref{FIG4}. One sees that for $s>30,$the
corresponding values of $\sigma_{s}$ are less than $10^{-5}$, and the
expansion (\ref{76}) could have been truncated at $s=30$ if an accuracy of
$1:10^{-5}$ had been sufficient.
The Chebyshev expansion coefficients $a_{j},\ j=1,2,..N$ of the wave function
$\bar{\psi},$ Eq.(\ref{L9a}) mapped into the interval $[-1,+1]$ are
illustrated in Fig. \ref{FIG6}.
The magnitude of these coefficients decreases exponentially with the index
$j$, reaching the value $\simeq10^{-14}$ for $j=100$ and beyond. Had the
expansion been truncated at $j=30,$ the error of the wave function would have
been $1:10^{-5}.$

In conclusion, the accuracy of the solution of Eq. (\ref{L-S}) with the
spectral expansion method for a nonlocality described by Eq. (\ref{K-SIMP}),
and using $N=51$ Chebyshev polynomials is $1:10^{-5}$, while using $N=301$
polynomials, the accuracy is better than $1:10^{-10}.$

\section{The Sturmian Expansion Method}

The method consists in expanding the unknown solution $\psi$ of Eq.
(\ref{L-S}) into a basis set of $N$ "global" functions, formulate an equation
for the expansion coefficients $a_{i}$, and if $N$ is not large enough,
iteratively correct for the truncation error. The expansion into Sturmian
functions was described in Ref. \cite{STURM-GR}, and only a few basic
equations relevant for the present case will be repeated here, while Appendix
B contains further details. References to many other applications of Sturmian
functions to the solution of physics problems can also be found in Ref.
\cite{STURM-GR}.

For the present application the operator $\mathcal{O}$ in the general
one-dimensional integral equation
\begin{equation}
\psi(r)=F(r)+\int_{0}^{\infty}\mathcal{O(}r,r^{\prime \prime})\psi
(r^{\prime \prime})\ dr^{\prime \prime}, \label{L1}%
\end{equation}
to be solved for $\psi$\ is, in view of Eq. (\ref{L-S}), given by
\begin{equation}
\mathcal{O(}r,r^{\prime \prime})=\int_{0}^{\infty}\mathcal{G}_{0}(r,r^{\prime
})K(r^{\prime}r^{\prime \prime})dr^{\prime}, \label{L10}%
\end{equation}
where it is assumed that the local part $V$ of the potential in Eq.
(\ref{L-S}) has been set to zero. The function $F$ is the driving term and
$\mathcal{G}_{0}$ is the Green's function, described in Eqs. (\ref{L7}) and
(\ref{L8}), respectively, both assumed to be known. The shorthand form of Eq.
(\ref{L1}) is%
\begin{equation}
\psi=F+\mathcal{O}\psi.\  \label{L2}%
\end{equation}
The iterative solution of Eq. (\ref{L1}) is achieved by approximating the
operator $\mathcal{O}$ by a separable representation $\mathcal{O}_{N}$ of rank
$N$, defining the remainder $\Delta_{N}^{(1)}$ as%
\begin{equation}
\Delta_{N}=\mathcal{O}-\mathcal{O}_{N}, \label{27}%
\end{equation}
and iterating on the remainder. \ 

The approximate discretization of the kernel $\mathcal{O}$ into a
representation of rank $N$ is accomplished by using a set of auxiliary
positive or negative energy sturmian functions $\Phi_{s}(r),$ $s=1,2,..,N$ and
is of the form
\begin{equation}
\mathcal{O}_{N}(r,r^{\prime})\  \mathcal{=}\sum_{s=1}^{N}\mathcal{O\ }\Phi
_{s}\rangle \frac{1}{\langle \Phi_{s}\bar{V}\Phi_{s}\rangle}\langle \Phi_{s}%
\bar{V}. \label{L4}%
\end{equation}
Here the symbol $\rangle$ denotes that the quantity to the left of it is
evaluated at position $r$, and $\langle$\ denotes that the quantity to the
right of it is evaluated at $r^{\prime}.$ The bra-ket $\langle \Phi_{s}\bar
{V}\Phi_{s}\rangle$ denotes the integration $\langle \Phi_{s}\bar{V}%
\Phi_{s^{\prime}}\rangle=\int_{0}^{\infty}\Phi_{s}(r)\bar{V}(r)\Phi
_{s^{\prime}}(r)dr$\ where $\langle \Phi_{s}$ is \emph{not} the complex
conjugate of $\Phi_{s},$ and $\bar{V}(r)$ is the local potential used in the
definition of the Sturmians.

The sturmian functions $\Phi_{s}$ are eigenfunctions of the integral kernel
$\mathcal{G}_{0}(r,r^{\prime})\bar{V}(r^{\prime})$%
\begin{equation}
\eta_{s}\Phi_{s}(r)=\int_{0}^{\infty}\mathcal{G}_{0}(r,r^{\prime})\bar
{V}(r^{\prime})\Phi_{s}(r^{\prime})dr^{\prime},\  \  \ s=1,2,3,.... \label{L5}%
\end{equation}
with $\eta_{s}$ the eigenvalue, and $\bar{V}(r^{\prime})$ the sturmian
potential. The differential Schr\"{o}dinger equation corresponding to Eq.
(\ref{L5}) is\
\begin{equation}
\mathcal{(}d^{2}/dr^{2}+E)\  \Phi_{s}=\Lambda_{s}\bar{V}\  \Phi_{s}, \label{8}%
\end{equation}
with $\Lambda_{s}=1/\eta_{s}$. The Sturmians for positive energies are not
square integrable, but they are orthogonal to each other with the weight
factor $\bar{V}$ (that is assumed to decrease sufficiently fast with $r)$. The
normalization of the Sturmians adopted for most of the present discussion is
\begin{equation}
\langle \Phi_{s}\bar{V}\Phi_{s^{\prime}}\rangle=\eta_{s}\delta_{s,s^{\prime}}.
\label{L12}%
\end{equation}
Because of the completeness of the sturmian functions, one has the identity
\begin{equation}
\delta(r-r^{\prime})=\sum_{s=1}^{\infty}\Phi_{s}(r)\frac{1}{\langle \Phi
_{s}\bar{V}\Phi_{s}\rangle}\Phi_{s}(r^{\prime})\bar{V}(r^{\prime}),
\label{L13}%
\end{equation}
which shows that the larger the number of terms $N$ in the expansion
(\ref{L4}), the better is the approximation of $\mathcal{O}_{N}$ to
$\mathcal{O}$, provided that $\mathcal{O}$ is compact.

The first step in the solution of Eq. (\ref{L2}) is to obtain the solution
$\mathcal{F}$ of the approximate equation
\begin{equation}
\mathcal{F}(r)=F(r)+\int_{0}^{\infty}\mathcal{O}_{N}\mathcal{(}r,r^{\prime
\prime})\mathcal{F}(r^{\prime \prime})\ dr^{\prime \prime}. \label{L14}%
\end{equation}
The solution can be obtained algebraically \cite{Canton} by \ making the
ansatz \cite{STURM-GR}
\begin{equation}
\mathcal{F}(r)=F(r)+\sum_{s=1}^{N}\ c_{s}\ |\mathcal{O\ }\Phi_{s}\rangle_{r},
\label{L15}%
\end{equation}
and the coefficients $c_{s}$ , $s=1,2,..N$, are obtained from the solution of
the matrix equation
\begin{equation}
\sum_{s^{\prime}=1}^{N}\left(  \delta_{s,s^{\prime}}-M_{s,s^{\prime}}\right)
c_{s^{\prime}}=\frac{1}{\langle \Phi_{s}\bar{V}\Phi_{s}\rangle}\langle \Phi
_{s}\bar{V}\ F\rangle, \label{L16}%
\end{equation}
where
\begin{equation}
M_{s,s^{\prime}}=\frac{1}{\langle \Phi_{s}\bar{V}\Phi_{s}\rangle}\langle
\Phi_{s}\bar{V}|\mathcal{O}\Phi_{s^{\prime}}\rangle=\langle \Phi_{s}%
K\Phi_{s^{\prime}}\rangle. \label{L17}%
\end{equation}
Eqs. (\ref{L16}) and (\ref{L17}) are obtained by inserting (\ref{L15}) into
(\ref{L14}), making use of Eq. (\ref{L4}), multiplying the resulting equation
by $[1/\langle \Phi_{s}\bar{V}\  \Phi_{s}\rangle]\  \langle \Phi_{s}(r)\  \bar
{V}(r)$ on both sides, integrating over $r$, and making use of the
normalization (\ref{L12}). The matrix element $\langle \Phi_{s}\bar
{V}|\mathcal{O}\Phi_{s^{\prime}}\rangle \ $involves a triple integral, while
the result $\langle \Phi_{s}K\Phi_{s^{\prime}}\rangle$ in Eq. (\ref{L17})
requires only a double integral in view of Eq. (\ref{L5}). This simplification
is one of the advantages of using Sturmian functions in the expansion
(\ref{L4}) of $\mathcal{O}.$ The integrals in the equations above are carried
out using a Gauss-Chebyshev procedure \cite{STRING} that has high accuracy,
and does not require the use of the $SVD$ decomposition.

Had the expansion $\sum_{s=1}^{N}$ $c_{s}|\mathcal{G}_{0}\mathcal{K}\Phi
_{s}\rangle$ in Eq. (\ref{L15}) been replaced by the expansion $\sum
_{s=1}^{\bar{N}}d_{s}|\Phi_{s}\rangle$, then the relation between the two sets
of coefficients would have been
\begin{equation}
d_{s}=\sum_{s^{\prime}=1}^{N}M_{s,s^{\prime}}c_{s^{\prime}}~~s=1,2,..\bar{N}.
\label{L18}%
\end{equation}
Since $\bar{N}$ can be larger than $N,$ the expansion in Eq. (\ref{L15}) is
preferable because $\mathcal{O\ }\Phi_{s}\rangle_{r}$ may lie outside of the
space spanned by the functions $\Phi_{s}\rangle_{r}$, and hence is more
general. However, for the present application $\bar{N}=N.$

\begin{figure}
[ptb]
\begin{center}
\includegraphics[
height=2.1352in,
width=2.8392in
]%
{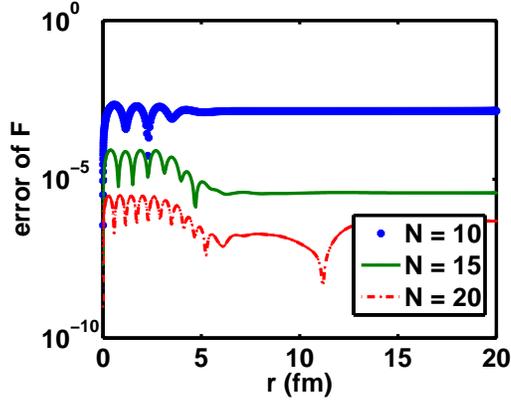}%
\caption{The error of the sturmian expansion of $\mathcal{F}$, the solution of
Eq. (\ref{L14}), for different numbers $N$ of Sturmian basis functions. The
auxiliary Sturmian potential $\bar{V},$ is of the Woods Saxon form, given by
Eq. (\ref{VPB}) with parameters ($V_{0},R,a)=5fm^{-2},11fm,0.5fm)$. The radial
interval is $0\leq r\leq20\ fm$, and the wave number is $k=0.5fm^{-1}$}%
\label{FIG9}%
\end{center}
\end{figure}

Results for the function $\mathcal{F},$ given by Eqs. (\ref{L15}) to
(\ref{L17}), were compared with the wave function obtained by the Spectral
Chebyshev method described in section III. The difference between the two
results was taken as a measure of the accuracy of the Sturmian expansion, that
in turn depends on the range $R$ of the Sturmian auxiliary potential $\bar
{V},$ and the number $N$ of sturmian functions used. Here $R$ is defined in
Eq. (\ref{VPB}). The dependence of the accuracy of $\mathcal{F}$ on the number
$N\ $\ of sturmians\ for a fixed value of \ $R=11\ fm$ is illustrated in Fig.
\ref{FIG9}. It shows that with $N=20$ an accuracy of $1:10^{-6}$ is achieved.
The accuracy for different values of $R$ and $N$ is is summarized in Table
\ref{TABLE2} and is illustrated in Figs. \ref{FIG12} and \ref{FIG13}. The
computational time, carried out in MATLAB, is listed in Table \ref{TABLE3}.
This result does not include the time required to calculate the Sturmian functions.%

\begin{table}[tbp] \centering
\begin{tabular}
[c]{|l|l|l|l|}\hline
$N_{S}$ & $N_{P}$ & $~~~~\  \mathcal{F}$ & $1^{st}iter$\\ \hline \hline
$10$ & $301$ & $2\times10^{-3}$ & $7\times10^{-5}$\\ \hline
$15$ & $301$ & $9\times10^{-5}$ & $2\times10^{-6}$\\ \hline
$20$ & $453$ & $3\times10^{-6}$ & $3\times10^{-7}$\\ \hline
\end{tabular}
\caption{Accuracy of the Sturmian expansion. \(N_S \) is the number of Sturmians, \(N_P\) is the number of support points in the radial interval [0,15 {\sl fm}]}\label{TABLE2}%
\end{table}%
\begin{table}[tbp] \centering
\begin{tabular}
[c]{|l|l|l|l|}\hline
$N_{S}$ & $N_{P}$ & $\mathcal{F}\ (s)$ & $1^{st}iter.(s)$\\ \hline \hline
$10$ & $301$ & $11$ & $0.04$\\ \hline
$15$ & $301$ & $11$ & $0.04$\\ \hline
$20$ & $453$ & $26$ & $0.11$\\ \hline
\end{tabular}
\caption{Computing time of the Sturmian expansion. \(N_S \) is the number of Sturmians, \(N_P\) is the number of support points in the radial interval [0,15 {\sl fm}]}\label{TABLE3}%
\end{table}%
.\medskip

\subsection{Iterative corrections to $\mathcal{F}.$}

The iterative correction to the Sturmian expansion of $\mathcal{F}$, Eq.
(\ref{L15}) is carried out by defining the remainder $\Delta_{N}$ according to
Eq. (\ref{27}) and iterating on the remainder. If the norm of $\Delta_{N}$ is
less than unity, the iterations should converge \cite{STURM-GR}. Since the
numerical complexity of performing iterations is less than the complexity of
solving a linear equation with a matrix of large dimension, this method can be
computationally advantageous, as is shown in the $4^{th}$column of Table
\ref{TABLE3}, and furthermore the exact eigenfunctions of the operator
$\mathcal{O}$ need not be known.\ 

The iterative corrections to $\psi$ proceed according to
\begin{equation}
\psi=\mathcal{F}+\chi_{1}+\chi_{2}+.... \label{L20}%
\end{equation}
where the $\chi_{n+1}$ are related to $\chi_{n}$ through the iterative
equation \cite{STURM-GR}%
\begin{equation}
\chi_{n+1}=\mathcal{O}_{N}\chi_{n+1}+\Delta_{N}\chi_{n},\  \ n=0,1,2,..,
\label{L21}%
\end{equation}
with $\chi_{0}=\mathcal{F}.$ If the expansion of $\chi_{n+1}$ is given by
\begin{equation}
\chi_{n+1}(r)=\sum_{s=1}^{N}d_{s}^{\ (n+1)\ }\Phi_{s}(r) \label{L18a}%
\end{equation}
then the coefficients $d_{s}^{\ (n+1)}$ obey the algebraic equation
\begin{equation}
\sum_{s^{\prime}=1}^{N}\left(  \delta_{s,s^{\prime}}-M_{s,s^{\prime}}\right)
d_{s^{\prime}}^{\ (n+1)}=\frac{1}{\langle \Phi_{s}\bar{V}\Phi_{s}\rangle
}\langle \Phi_{s}\bar{V}\  \Delta_{N}\chi_{n}\rangle, \label{L22}%
\end{equation}
where $M_{s,s^{\prime}}$ is given by Eq. (\ref{L17}). However, before
inserting the numerical value of $\chi_{n}(r)$ into the right hand side of Eq.
(\ref{L22}), Eq. (\ref{L18a}) is replaced by $\sum_{1}^{N}$ $c_{s}%
^{(n)}|\mathcal{G}_{0}\mathcal{K}\Phi_{s}\rangle$, with $c_{s}^{(n)}%
=\sum_{s^{\prime}=1}^{N}(M^{-1})_{s,s^{\prime}}d_{s^{\prime}}^{\ (n)}$,
according to Eq. (\ref{L18}).

\begin{figure}
[ptb]
\begin{center}
\includegraphics[
height=2.1672in,
width=2.8807in
]%
{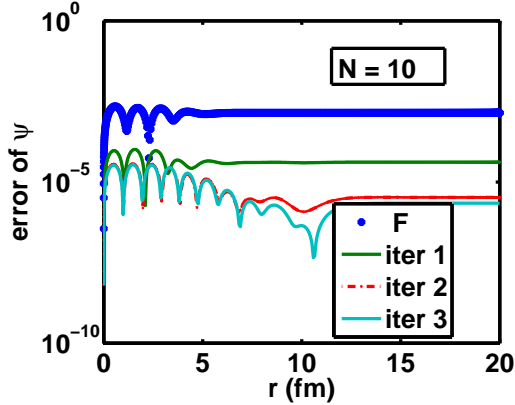}%
\caption{Iterative corrections to the wave function $\psi$, using 10
Sturmians, which are described in Fig. \ref{FIG9}. The result labeled $F$
illustrates the value of $\mathcal{F}$ as described in Fig. \ref{FIG9}. The
other results are obtained after 1, 2 or 3 iterations. The maxima of the
\ three curves occur at $3\times10^{-3}$, $\ 10^{-4}$, and $4\times10^{-5}$,
respectively.}%
\label{FIG10}%
\end{center}
\end{figure}

The results for the number of Sturmians $N=10$ is illustrated in Fig.
\ref{FIG10} for $R=11\ fm$. It is seen that as the number of iterations
increases, the result for $\psi$ converges but not, within the accuracy of the
calculation, to the value obtained from the spectral method. \ This is in
contrast with the case of a local potential. There the converged result of the
iterative Sturmian method was found to be\ in good agreement with the spectral
result \cite{STURM-GR} of $1:10^{-8}$, as compared with the less accurate
result for the present nonlocal case. It is suspected that the reason for this
difference is related to a difference in the treatment of the long-range part
of the potential, as is discussed further in Appendix $B$. On the other hand
the error of $\mathcal{F}$ (compared to $\psi)$ is much smaller in the present
nonlocal case than it is for the local case \cite{STURM-GR}. That difference
is very likely due to the presence of a repulsive core in the scattering
potential for the local case, but is absent in the nonlocal case.%

\begin{figure}
[ptb]
\begin{center}
\includegraphics[
height=2.0928in,
width=2.7838in
]%
{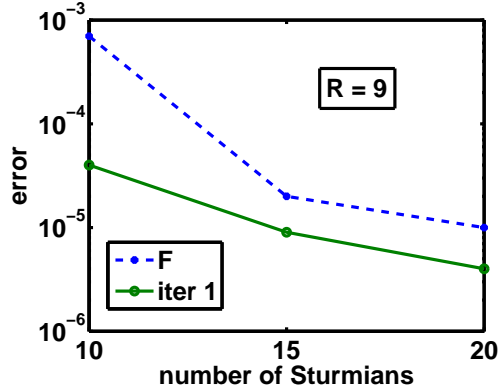}%
\caption{Maximum absolute error of $\psi$, as a function of the number of
Sturmian basis functions. The Sturmian potential $\bar{V}$ is defined in Eq.
(\ref{VPB}) with the parameters ($V_{0},R,a)$ given by ($-5\ fm^{-2}%
,\ 9\ fm,\ 0.5\ fm).$ The line labed $"F"$ represents the result for
$\mathcal{F}$, Eq. (\ref{L15}), while the line labeld "$iter\ 1"$ is obtained
by correcting $\mathcal{F}$ by one iteration, as described in the text.\ }%
\label{FIG12}%
\end{center}
\end{figure}
\begin{figure}
[ptb]
\begin{center}
\includegraphics[
height=2.0989in,
width=2.7908in
]%
{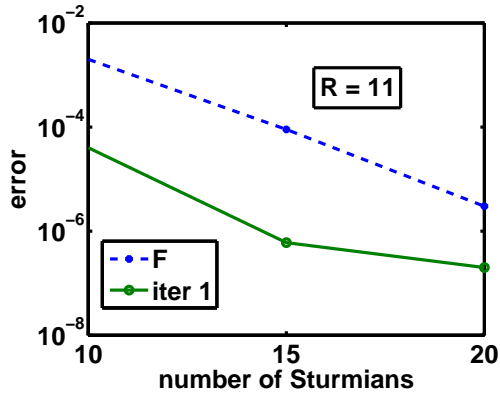}%
\caption{Same as for Fig. \ref{FIG12} for a larger range $R=11\ fm$ of the
sturmian potential.}%
\label{FIG13}%
\end{center}
\end{figure}

The effect of the range $R$ of $\bar{V}$ and the number $N$ of Sturmians is
displayed in Figs. \ref{FIG12} and \ref{FIG13}. These figures show that for a
given $N,$ the accuracy of $\mathcal{F}$ is better for a smaller value of $R,$
but the iterations give a more accurate value for $\psi$ for a larger value of
$R.$ Table \ref{TABLE3} shows that the iteration time is considerably less
than the time to compute $\mathcal{F}$, and in order to obtain the same
accuracy, the latter is considerably longer than the computing time for the
Spectral Chebyshev expansion.

\section{Summary and conclusions}

The solution of a radial Schr\"{o}dinger equation\ for $\psi(r)$ containing a
general nonlocal potential of the form $\int K(r,r^{\prime})\  \psi(r^{\prime
})\ dr^{\prime}$ is obtained by means of a spectral expansion into Chebyshev
polynomials \cite{IEM}, combined with a Singular Value Decomposition. For a
semi-separable kernel $K$ of rank one, that occurs in exchange scattering due
to the Pauli exclusion principle, a spectral expansion method has been
previously devised \cite{EXCH}. But for a general $K$ the present method is
quite different, and has not been presented before. The results for a
numerical example of $K$ given by a Perey-Buck ansatz \cite{Perey} are
calculated and their accuracy, as well as the required computer time, is
investigated as a function of the number of Chebyshev polynomials employed
(see Table \ref{TABLE1}). For a wave number $k=0.5\ fm^{-1}$ an accuracy of
between $1:10^{-5}$ to $1:10^{-10}$ is obtained as the number of polynomials
in the whole radial interval is increased from $50$ to $300$. A second method
is presented, based on an expansion into $N$ Sturmian functions of positive
energy, supplemented by an iteration procedure \cite{STURM-GR}. For $N=15$ an
accuracy of $1:10^{-4}$ is obtained without iterations, and after one
iteration the accuracy is increased to $1:10^{-6}$, as shown in Fig.
\ref{FIG13}$.$ The iterations converge quickly, but not to exactly to the
value given by the spectral method. It is suspected that the reason is due to
the range of the Sturmian functions employed not being as large as the range
of the nonlocal potential $K,$ as is further discussed in Appendix $B.$

The method, being applicable to a general nonlocality $K$, opens the way to
formulate optical potentials that incorporate the physical effects that are
the source of the nonlocality. This study is particularly relevant for
Astrophysics, where the colliding nuclei are generally unstable, and hence the
resulting nonlocalities are more pronounced.

Acknowledgment:\ The author is grateful to Professor M. Jaghoub for rekindling
his interest into the subject of nonlocalities.

\bigskip

{\LARGE Appendix A: The rank of a matrix}

Various definitions of the rank of a matrix can be found in Chapter 9, p. 432
of Ref. \cite{McQ}. According to one of \ the definitions, the rank of a
matrix $A$ is the order of the largest square sub-matrix of $A$ whose
determinant in not equal to zero. A practical definition is obtained via the
SVD decomposition of the matrix $A\ $according to which the rank of $A$ is the
number of non-zero singular values of $A.$ For example: given two column
vectors $\vec{u}$ and $\vec{v},$ both having $N$ elements $u_{i}$ and $v_{i}$,
$i=1,2,..N,$ then $v^{T}\cdot u=\sum_{i=1}^{N}v_{i}u_{i}$ is a number, while
$u\ v^{T}$ is a $N\times N$ matrix $W.$ Here the superscript $T$ means
transposition. The matrix $W$ has only one nonvanishing singular value, hence
its rank is $1.$ The $4\times4$ matrix $A$ given on p. 433 of Ref. \cite{McQ}
has $3$ non zero singular values, hence its rank is $3,$ in agreement with
Ref. \cite{McQ}.

In the case of exchange scattering, the non locality $K$ is given by
$K(r,r^{\prime})=f(r)g(r^{\prime})$ for $r\prime<r$, and $g(r)f(r^{\prime})$
for $r^{\prime}\geq r$.\ Once the radial distance $r$ is discretized into $N$
values $r_{i}$, the function of $r_{i}^{\prime}$ is interpreted as a line
vector, while the function of $r_{i}$ is taken as a column vector, and the
kernel $K$ is the separable combination of two rank $1$ matrices, each of
dimension $N\times N$, and hence it is called semi-separable of rank
one.\bigskip

{\LARGE Appendix B: Sturmian functions}

\begin{figure}
[ptb]
\begin{center}
\includegraphics[
height=2.1352in,
width=2.8392in
]%
{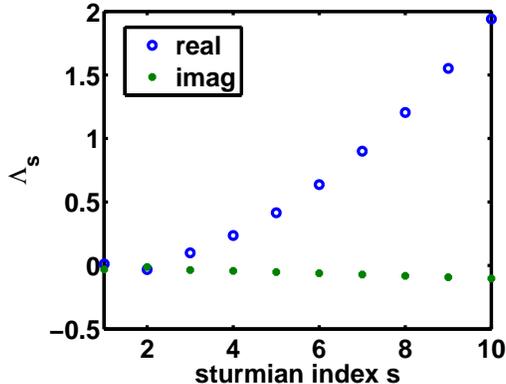}%
\caption{The Sturmians eigenvalues\ $\Lambda_{s}$ of Eq. (\ref{8}) obtained
with a Woods--Saxon potential $V_{WS}$ described in the caption of Fig.
\ref{FIG12}, with a wave number $k=0.5\ fm^{-1}$ }%
\label{FIG14}%
\end{center}
\end{figure}

Some properties of Sturmian functions \cite{STURM-GR} will be recalled here.
According to Eq. (\ref{L5}) Sturmian functions are defined in the radial
interval $[0,\infty]$. However, the numerical evaluation of (\ref{L5}) has to
be carried out only in the interval $0\leq r\leq R_{\max}$, beyond which the
Sturmian potential is negligible. An example of the spectrum of the $\Lambda$
eigenvalues for the Woods-Saxon potential $V_{WS}$ illustrated in Fig.
\ref{FIG3}, for a wave number $k=0.5\ fm^{-1}$, is shown in Fig. \ref{FIG14}.
The imaginary parts are slightly negative, while the real parts are
monotonically positive. This is in contrast with the case of a sturmian
potential that has a repulsive core, for which some of the eigenvalues have a
negative real part, and a correspondingly positive imaginary part
\cite{STURM-GR}.%

\begin{figure}
[ptb]
\begin{center}
\includegraphics[
height=2.2191in,
width=2.9507in
]%
{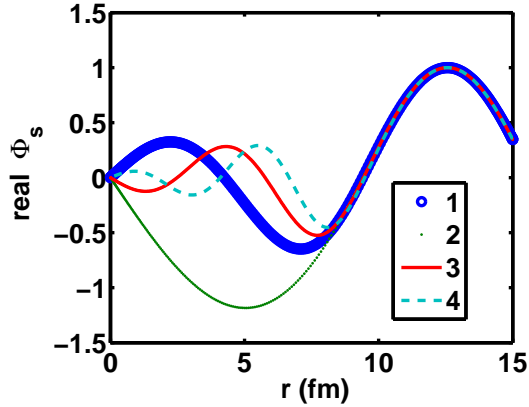}%
\caption{The real parts of Sturmian eigenfuncions of Eq.\ (\ref{8}) obtained
with a Woods--Saxon potential $V_{WS}$ described in the caption to Fig.
\ref{FIG12}, whose eigenvalues are illustrated in Fig. \ref{FIG14}}%
\label{FIG15}%
\end{center}
\end{figure}

\begin{figure}
[ptb]
\begin{center}
\includegraphics[
height=2.2347in,
width=2.9715in
]%
{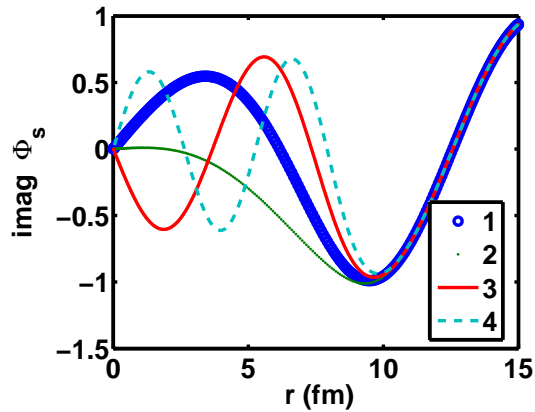}%
\caption{Same as for Fig. \ref{FIG15}, for the imaginary parts of the Sturmian
functions.}%
\label{FIG16}%
\end{center}
\end{figure}

The radial $r-$dependence of some of the Sturmian functions is illustrated in
Figs. \ref{FIG15} and \ref{FIG16}. Their normalizations differ from Eq.
(\ref{L12}) so that all acquire the same Hankel function asymptotic behavior,
since they obey Eq. (\ref{L5}). The Sturmian potential $\bar{V}$ has a range
$R=9\ fm,$ and the figures show that for $r>8\ fm$, these Sturmian functions
are not strongly linearly independent from each other, and hence cease to form
a practical complete expansion set. This is the reason why the Sturmian
expansion for a Woods-Saxon potential with a range $R=9\ fm$ does not give a
result as precise as the expansions for $R=11\ fm$%
and even that range may not suffice to obtain an accuracy better than
$1:10^{7},$ displayed in Fig. \ref{FIG13}. This conclusion is further
corroborated by examining the magnitude of the kernel $K$ in the region where
the Sturmian functions loose most of their independence, which is of the order
of $10^{-7}\ fm^{-2}.$

\end{document}